\documentclass[twocolumn,aps,showpacs,superscriptaddress,pre]{revtex4}

\usepackage{amsmath,amssymb,graphicx}
\usepackage{bm}
\usepackage{algorithmic}
\usepackage{enumerate}
\usepackage{color}
\usepackage{stmaryrd}
\usepackage{times}

\definecolor{yblue}{rgb}{0.06, 0.3, 0.57}
\usepackage[pdftex]{hyperref}
\hypersetup{colorlinks=true,linkcolor=yblue,citecolor=yblue,urlcolor=yblue}

\begin{document}

\title{Numerical simulations of Ising spin glasses with free boundary conditions: the role of droplet excitations and domain walls}

\author{Wenlong Wang}
\email{wenlongcmp@gmail.com}
\affiliation{Department of Physics and Astronomy, Texas A$\&M$ University, 
College Station, Texas 77843-4242, USA}

%\author{Helmut G. Katzgraber}
%\affiliation{Department of Physics and Astronomy, Texas A$\&M$ University, 
%College Station, Texas 77843-4242, USA}
%\affiliation{Santa Fe Institute, 1399 Hyde Park Road, Santa Fe, New Mexico 
%87501, USA}

\begin{abstract}

The relative importance of the contributions of droplet excitations and domain walls on the ordering of short-range Edwards-Anderson spin glasses in three and four dimensions is studied. We compare the overlap distributions of periodic and free boundary conditions using population annealing Monte Carlo. For system sizes up to about 1000 spins, spin glasses show non-trivial spin overlap distributions. Periodic boundary conditions can trap diffusive domain walls which can contribute to small spin overlaps, and the other contribution is the existence of low-energy droplet excitations within the system. We use free boundary conditions to minimize domain-wall effects, and show that low-energy droplet excitations are the major contribution to small overlaps in numerical simulations. Free boundary conditions has stronger finite-size effects, and is likely to have the same thermodynamic limit with periodic boundary conditions.

\end{abstract}

\pacs{75.50.Lk, 75.40.Mg, 05.50.+q, 64.60.-i}
\maketitle

\section{Introduction}

The nature of the ordering of short-range Edwards-Anderson (EA) spin glasses \cite{edwards:75} is a subject of long-standing controversy \cite{mezard:84,moore:98,drossel:00,krzakala:00,palassini:99,palassini:00,marinari:00,marinari:00a,marinari:00c,katzgraber:01,middleton:01,hatano:02,katzgraber:02,katzgraber:03,katzgraber:03f,hed:07,leuzzi:08,alvarez:10a,yucesoy:12,billoire:13,yucesoy:13b,wang:14,wang:15a,wang:16a,Timo:16a}. The infinite-range Sherrington-Kirkpatrick (SK) model \cite{sherrington:75} is known to have an infinite number of pure states, described by replica symmetry breaking \cite{parisi:79,parisi:80,parisi:83}. In the context of short-range spin glasses, there are two similar and plausible ways to have many pairs of pure states, but in terms of metastates \cite{aizenman:90,newman:96a,newman:97}.
For a finite large volume of spins, there might be one pair of pure states present, the chaotic pairs picture \cite{newman:92,newman:97} or many pairs of pure states, the non-standard replica symmetry breaking (RSB) picture \cite{newman:97,read:14a}, both with chaotic size dependence and space-filling domain walls. The droplet picture on the other hand,
developed by McMillan \cite{mcmillan:84b}, Bray and
Moore~\cite{bray:86}, as well as Fisher and
Huse~\cite{fisher:86,fisher:87,fisher:88}, is an example of the simple
scenario that there is only a single pair of pure states
and the thermodynamic limit is defined in the usual way. In the droplet/scaling picture, domain walls are fractal surfaces, not space-filling.

Many numerical simulations have been conducted to study the ordering of the EA model \cite{mezard:84,moore:98,drossel:00,krzakala:00,palassini:00,marinari:00,marinari:00a,marinari:00c,katzgraber:01,middleton:01,hatano:02,katzgraber:02,katzgraber:03,katzgraber:03f,hed:07,leuzzi:08,alvarez:10a,yucesoy:12,billoire:13,yucesoy:13b,wang:14,Timo:16a}
with a confusing mixture of results, in particular whether domain walls are space-filling, and there is a finite weight near zero overlap in the spin overlap distribution function $P(q)$. According to the droplet/scaling picture, the free energy cost to flip a droplet of size $\ell$ scales as $\ell^{\theta}$, where $\theta>0$ is a stiffness exponent, which is expected to be the same for domain walls and droplet excitations. On the other hand, RSB predicts that $\theta=0$ for droplet excitations. Consequently $P(0)$ scales as $\ell^{-\theta}$. Therefore, if $P(0)$ is finite, this means there are large-scale excitations in the system with $O(1)$ cost in free energy. Otherwise, there is a unique ordering of spins without system-size excitations. When a domain wall is introduced, there are $\ell^{d_s}$ spins on the surface of the domain wall, where $d_s$ is the fractal dimension of the domain wall. In the droplet/scaling picture, the surface is a fractal with $\rm{D-1} \leq d_s \leq \rm{D}$, while in RSB the surface is space-filling with $d_s=\rm{D}$, in D dimensions.
%In numerical simulations, the two exponents are mainly measured by studying the scaling of domain-wall free energies, energy or entropy.
To leading order without finite-size corrections, domain walls appear to be fractals and the weights near zero overlap is finite \cite{palassini:00,marinari:00,katzgraber:01,katzgraber:02} for the system sizes currently accessible. New statistics or finite-size corrections are therefore intensively developed, and point to different scenarios \cite{yucesoy:12,middleton:13,monthus:13,wang:14,wittmann:14a}. 

In this work, we focus on the weights near zero overlap $P(0)$.
\textit{We are interested in the question: if the droplet/scaling picture holds, could it be that $P(0)$ is a finite constant trivially because of trapped diffusive domain walls in the usually applied periodic boundary conditions (PBC), or low-energy droplet excitations are the dominate contribution?} Note that boundary conditions is only relevant to the EA model, not the SK model. It is not possible to tell this apart using periodic boundary conditions. Thermal boundary conditions (TBC) was used to reduce domain-wall effects, answered this question to some extent and indicated the answer is perhaps negative. The answer however is not completely clear, because TBC limits fluctuations only between periodic and anti-periodic boundary conditions according to the Boltzmann weights in each spatial direction, can still trap domain walls, and also overlaps between different boundary conditions are introduced. In this work, we minimize the domain-wall effects using free boundary conditions (FBC). FBC is probably the best boundary condition one can work with to separate the two effects. It is fruitful to compare PBC and FBC to answer this question clearly. \textit{Our strategy is as follows: (1) If domain-wall effects dominate, FBC should have stronger ordering than PBC, and (2) If droplet excitations dominate, FBC should make the ordering weaker or no change, for finite systems.} In this context, a stronger ordering means a smaller $P(0)$ and a weaker ordering means a larger $P(0)$. Our results show droplet excitations dominate $P(0)$, in line with that of TBC. The comparison of the PBC and FBC overlap functions also motivated us to draw a conclusion that PBC and FBC are likely to have the same thermodynamic limit, which is crucial in interpreting the FBC data properly.

It is however also well known that FBC introduces new finite-size effects as a substantial fraction of the spins are on the surface. Therefore, while useful when compared with PBC, FBC generally has stronger finite-size effects which could be misleading when looking for a trend with limited system sizes. FBC was used in the early work of Ref.~\cite{katzgraber:02} in revealing the nature of ordering of short-range spin glasses, and results of small system sizes were reported. In this work, we conduct large-scale Monte Carlo simulations, focus on the behaviour of $P(0)$ as a function of the system size, and compare PBC and FBC in both three and four dimensions. The existence of low-energy droplet excitations, and whether droplet excitations and domain walls have the same stiffness exponent have also been intensively studied. In Refs.~\cite{palassini:00,hartmann:02}, a small perturbation is added to the Hamiltonian such that the ground state energy increases more than the excited states to detect changes in the ground state, and hence the existence of low-energy droplet excitations. In Ref.~\cite{hartmann:04a}, various form of droplet excitations are generated and the stiffness exponents were studied in two dimensions.

The paper is organized as follows. We first discuss the model, observables and simulation methods in
Sec.~\ref{model}, followed by numerical results in Sec.~\ref{results}. Concluding remarks are stated in Sec.~\ref{summary}.

\section{Model, observables and methods}
\label{model}

We study the three-dimensional (3D) and four-dimensional (4D) Edwards-Anderson Ising spin-glass model \cite{edwards:75} defined by the Hamiltonian
\begin{equation}
H = - \sum_{\langle ij \rangle} J_{ij} S_i S_j ,
\label{eq:ham}
\end{equation}
where $S_i=\pm 1$ are Ising spins and the sum is over nearest neighbors
on a hyper-cubic lattice of linear size $L$ with number of spins $N=L^{\rm{D}}$. The random couplings $J_{ij}$
are chosen from a Gaussian distribution with mean $0$ and variance $1$.
A set of couplings $\{ J_{ij} \}$ therefore defines a disorder realization. We study free boundary conditions, as well as periodic boundary conditions. Our simulation is carried out using population annealing Monte Carlo \cite{hukushima:03,zhou:10,machta:10,wang:15e}. The simulation parameters are summarized in Table~\ref{table}. Note that the transition temperatures are $T_C \approx 1$ in 3D \cite{katzgraber:06} and $T_C \approx 1.8$ in 4D \cite{parisi:96}.

We study the spin overlap $q$ defined as
\begin{equation}
q=\dfrac{1}{N}\sum\limits_i S_{i}^{(1)} S_{i}^{(2)},
\label{e2}
\end{equation}
where spin configurations ``(1)'' and ``(2)'' are chosen independently
from the Boltzmann distribution, and its statistic $I(q_0)$
\begin{equation}
I(q_0)=\int_{-q_0}^{q_0} P(q) dq.
\label{e3}
\end{equation}
We study $I(0.2)$ unless otherwise specified.

%We use population annealing Monte Carlo (PAMC)
%\cite{hukushima:03,zhou:10,machta:10,wang:15e} to carry out the
%simulations. In PAMC, a large population of $R_0$ replicas of the
%system, each with the same disorder realization, are annealed in
%parallel from infinite temperature to a low target temperature, $T_0 =
%1/\beta_0$. The annealing schedule consists of a sequence of $N_T$
%temperatures equally spaced in inverse temperature $\beta$. In a
%temperature step from $\beta$ to $\beta^\prime$ the population is
%resampled; some members of the population are eliminated and some are
%reproduced. The mean number of copies of replica $i$ is proportional to
%the re-weighting factor, $\exp[-(\beta^\prime-\beta) E_i]$. The constant of proportionality is chosen so that the population size remains close to $R_0$. Following each resampling step, the population at $\beta'$ is acted on by $N_S = 10$ sweeps of the Metropolis algorithm. We simulate $M$ disorder realizations and measure overlaps at $T = T_0 = 0.2$ and $T=0.42$. Both of these temperatures are deep in the low-temperature spin-glass phase and should therefore not be affected by critical fluctuations.  Overlaps are then measured by pairing independent replicas in the population. The simulation parameters are shown in Table~\ref{table}.

\begin{table}
\caption{
Simulation parameters for the three-dimensional and four-dimensional EA models using population annealing Monte Carlo. D is the dimension of the system, BC is the boundary condition, $L$ is the linear system size, $R_0$ is the population size, $T_0$ is the
lowest temperature simulated, $N_T$ is the number of temperatures used in the annealing schedule, and $M$ is the number of disorder realizations studied. $N_S=10$ sweeps are applied to each replica at each temperature.
\label{table}
}
\begin{tabular*}{\columnwidth}{@{\extracolsep{\fill}} c c c c c c c}
\hline
\hline
D &BC &$L$  & $R_0$ & $T_0$ & $N_T$ & $M$ \\
\hline
3 &FBC &$4$  & $5\,10^4$ & $0.20$     & $101$   & $5000$ \\
3 &FBC &$6$  & $2\,10^5$ & $0.20$     & $101$   & $5000$ \\
3 &FBC &$8$  & $5\,10^5$ & $0.20$     & $201$   & $5000$ \\
3 &FBC &$10$ & $   10^6$ & $0.20$     & $301$   & $5000$ \\
3 &FBC &$12$ & $   10^6$ & $0.33$     & $301$   & $5000$ \\
4 &FBC &$3$  & $2\,10^4$ & $0.36$     & $101$   & $5000$ \\
4 &FBC &$4$  & $5\,10^4$ & $0.36$     & $101$   & $5000$ \\
4 &FBC &$5$  & $   10^5$ & $0.36$     & $101$   & $5000$ \\
4 &FBC &$6$  & $2\,10^5$ & $0.36$     & $201$   & $5000$ \\
4 &FBC &$7$  & $5\,10^5$ & $0.36$     & $201$   & $4400$ \\
4 &PBC &$8$  & $8\,10^5$ & $0.72$     & $301$   & $2000$ \\
4 &PBC &$3$  & $2\,10^4$ & $0.36$     & $101$   & $3000$ \\
4 &PBC &$4$  & $5\,10^4$ & $0.36$     & $101$   & $3000$ \\
4 &PBC &$5$  & $   10^5$ & $0.36$     & $101$   & $3000$ \\
4 &PBC &$6$  & $2\,10^5$ & $0.36$     & $201$   & $3000$ \\
4 &PBC &$7$  & $5\,10^5$ & $0.36$     & $201$   & $3000$ \\
4 &PBC &$8$  & $8\,10^5$ & $0.72$     & $301$   & $3000$ \\
\hline
\hline
\end{tabular*}
\end{table}

\section{Results}
\label{results}

In this section, we present our numerical results. We discuss the 3D results in Sec.~\ref{3D} and the 4D results in Sec.~\ref{4D}.

\subsection{The three dimensions}
\label{3D}

\begin{figure}[htb]
\begin{center}
\includegraphics[width=\columnwidth]{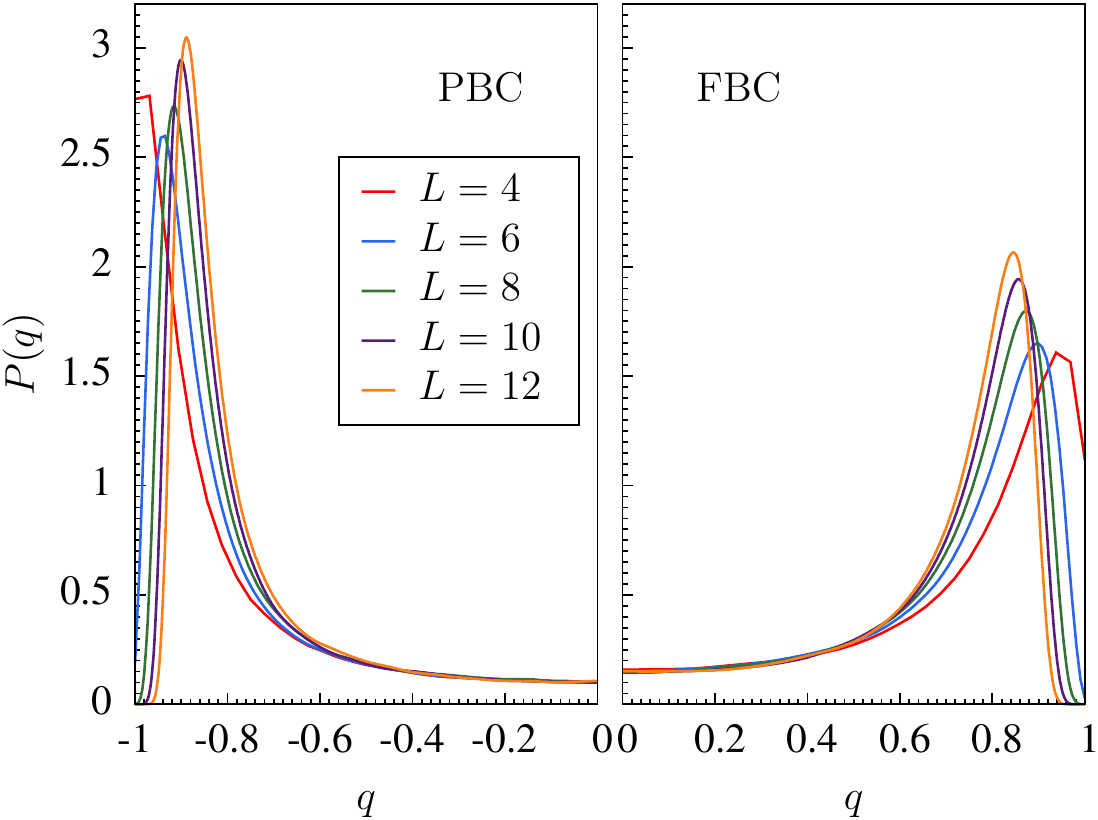}
\caption{(Color online)
Disorder-averaged spin overlap distributions $P(q)$ in 3D for
sizes $L=4, 6, 8, 10$, and $12$ at $T=0.42$ with periodic (left) and free (right) boundary conditions. The finite-size values of $\pm q_{\rm{EA}}$ decreases with system size. Note that FBC is less ordered than PBC for the system sizes studied.
}
\label{Pq3D}
\end{center}
\end{figure}

The disorder-averaged spin overlap distributions $P(q)$ for periodic and free boundary conditions are shown in Fig.~\ref{Pq3D}. The data for PBC is taken from a previous study of Ref.~\cite{wang:14}. Both display peaks at
finite-size values of $\pm q_{\rm EA}$, with the Edwards-Anderson order parameter $q_{\rm EA}$ decreases with $L$. For PBC at small $q$ the distribution is nearly independent of $L$, consistent with many past studies \cite{palassini:00,katzgraber:01,katzgraber:02,yucesoy:12}.

\begin{figure}[htb]
\begin{center}
\includegraphics[width=\columnwidth]{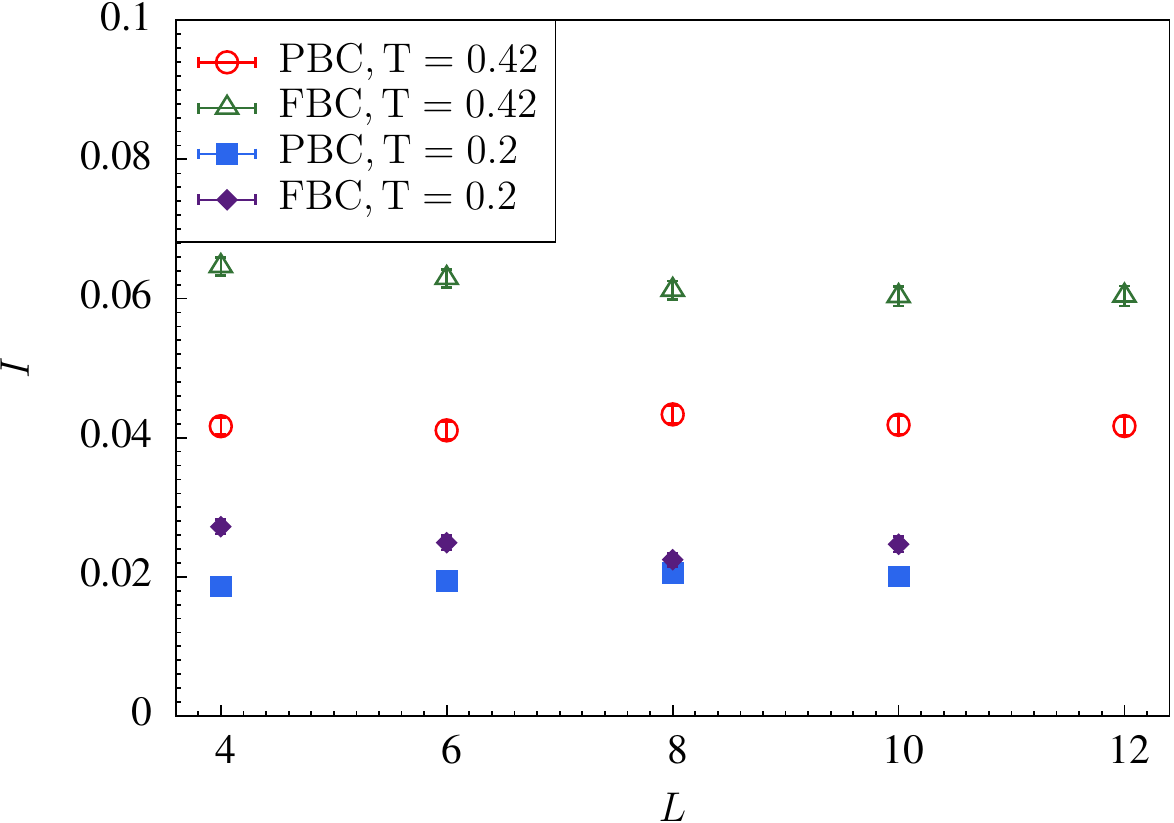}
\caption{(Color online)
$I$ as a function of system size $L$ in 3D for periodic and free boundary conditions. $I$ is approximately a constant for PBC, and is a fast decreasing function at small $L$ for FBC, but appears to level off and is bounded by the values of the PBC when $L$ increases.
}
\label{I3}
\end{center}
\end{figure}

It is worth noting in Fig.~\ref{Pq3D} that FBC ordering gets \textit{weaker} rather than stronger compared with PBC, suggesting that trapped domain walls in PBC cannot be used to argue why $P(0)$ is finite, and therefore there must be low-energy droplet excitations for the system sizes studied. To study the thermodynamic behaviour, we look at the statistic $I$ as a function of system sizes as shown in Fig.~\ref{I3}. For very small system sizes, $I$ appears to decrease with system size, similar to what was found in Ref.~\cite{katzgraber:02}. However, as system size gets larger, this trend does not appear to hold, especially at the lower temperature $T=0.2$, PBC appears to provide a lower bound for FBC. The same appears to hold in 4D, as shown in the next section.

It is easy to understand why $I$ is larger for small system sizes in FBC than PBC if droplet excitations dominate. If droplet/scaling picture holds, larger droplets can be excited by taking advantage of the free bonds on the surface. But $I$ would eventually become trivially the same and become 0 when system size gets larger, as the free energy cost inside the system would dominate, and the free bonds on the surface will not help.
If on the other hand RSB is correct, we again expect the excitations can take advantage of the free bonds on the surface, but expect this effect to be increasingly less important for larger system sizes. This would naturally suggest the same thermodynamic limit for FBC and PBC. Furthermore, the insensitivity of metastates to boundary conditions also support the scenario that FBC and PBC have the same thermodynamic limit. Therefore, we believe that the PBC $I$ is not just a lower bound for FBC, but the two would become the same in the thermodynamic limit. Our numerical results appear to support this conjecture, especially in 4D and lower temperatures, where finite-size effects are smaller.

\subsection{The four dimensions}
\label{4D}

\begin{figure}[htb]
\begin{center}
\includegraphics[width=\columnwidth]{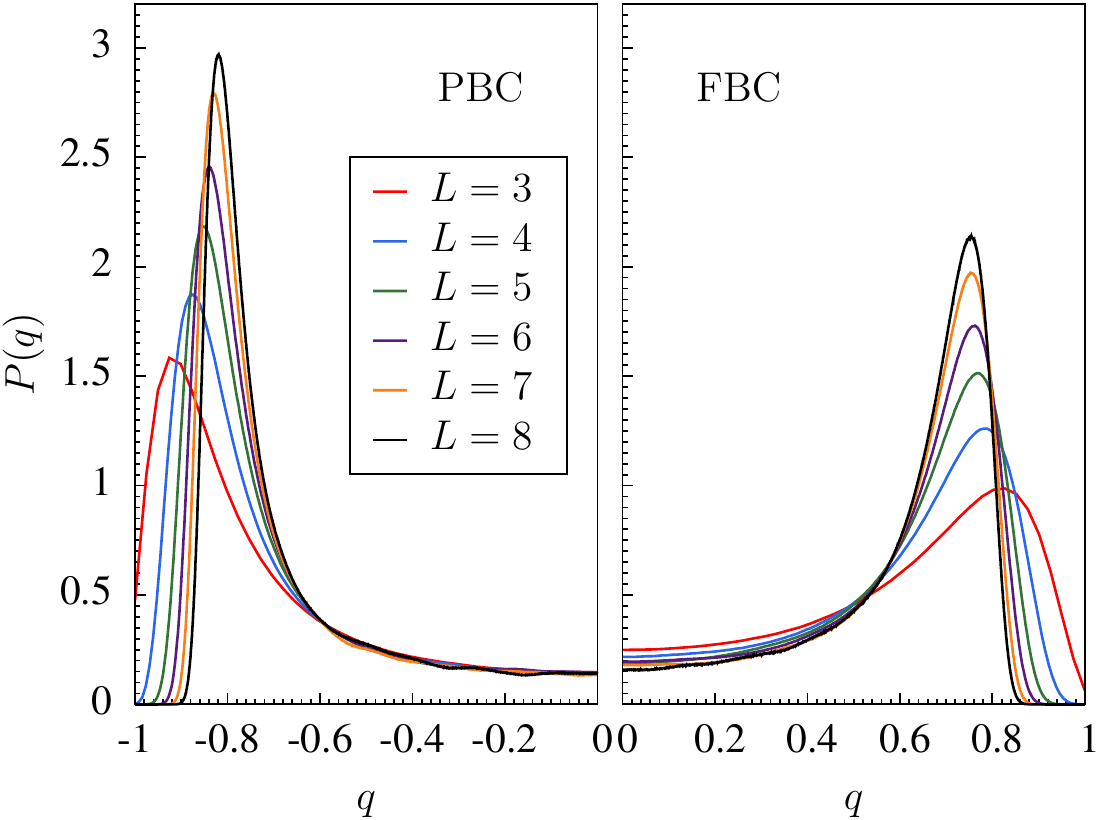}
\caption{(Color online)
Disorder-averaged spin overlap distributions $P(q)$ in 4D for
sizes $L=3, 4, 5, 6, 7$, and $8$ at $T=0.72$ with periodic (left) and free (right) boundary conditions. The finite-size values of $\pm q_{\rm{EA}}$ decreases with system size. Note that FBC is less ordered than PBC for the system sizes studied.
}
\label{Pq4D}
\end{center}
\end{figure}

\begin{figure}[htb]
\begin{center}
\includegraphics[width=\columnwidth]{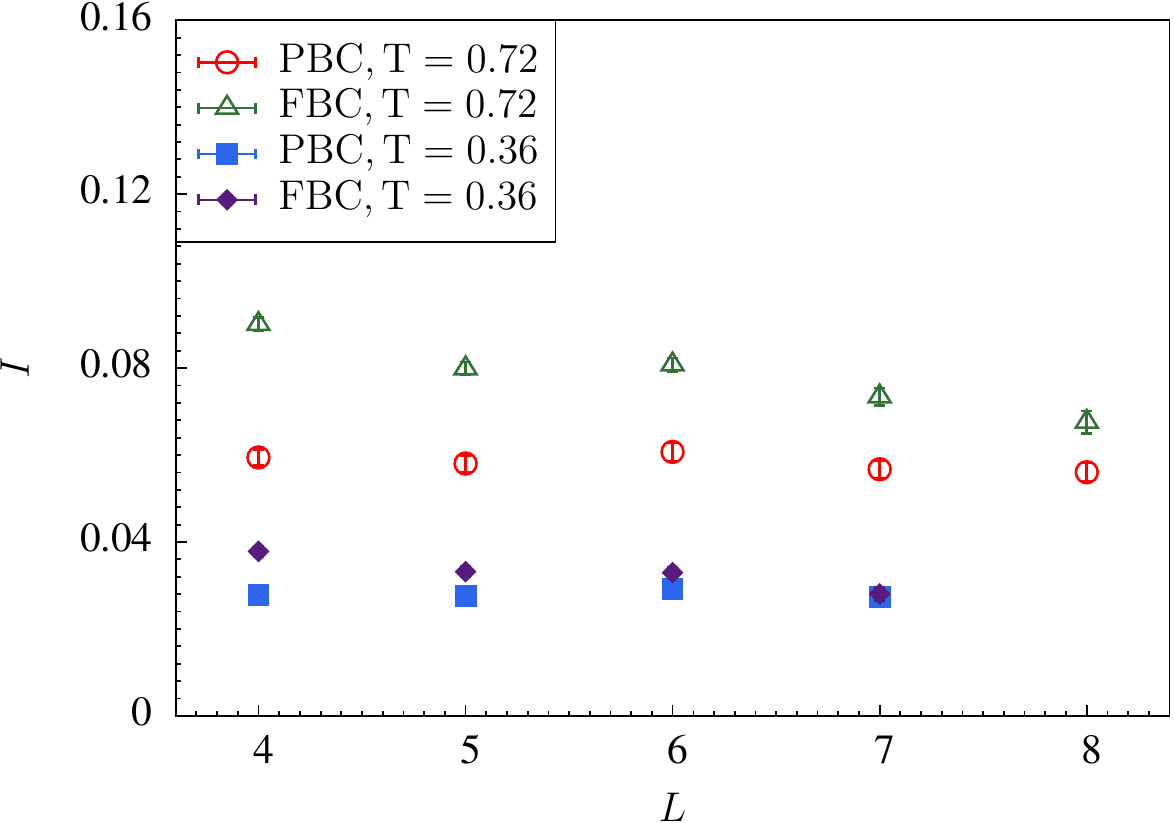}
\caption{(Color online)
$I$ as a function of system size $L$ in 4D for periodic and free boundary conditions. $I$ is approximately a constant for PBC, and is a fast decreasing function at small $L$ for FBC, but appears to level off and is bounded by the values of the PBC when $L$ increases. The fluctuation at $L=5$ and 6 is likely due to even-odd effects for small system sizes.
}
\label{I4}
\end{center}
\end{figure}

The disorder-averaged spin overlap distribution $P(q)$ and the statistic $I$ for PBC and FBC are shown in Figs.~\ref{Pq4D} and \ref{I4}, respectively. Similar behaviour as in 3D was found except that the trend becomes more apparent. By looking at $I$ of FBC alone, one may would like to argue $I$ is a decreasing function of $L$. However, we believe this is due the the strong finite-size effects of FBC. Note that in 3D, $I$ is also a decreasing function of $L$ up to around $L \approx 8$, and only appears to decrease slower or level off thereafter. We expect $I$ of FBC is still bounded by that of PBC in 4D.

If we believe FBC and PBC behave the same in the thermodynamic limit, then the controversy of whether $I \rightarrow 0$ when $L \rightarrow \infty$ continues, and it is perhaps indeed better to use PBC in numerical simulations which has smaller finite-size effects. Nevertheless, the comparison of FBC and PBC is interesting, and we think there is a clear conclusion that droplet excitations dominate $P(0)$ rather than trapped diffusive domain walls.

\section{Summary}
\label{summary}

We have investigated the spin overlap distributions of periodic and free boundary conditions. We find that the weights of small spin overlaps $I$ of FBC is larger and is bounded by that of PBC for finite systems. We conclude that droplet excitations is the major contribution to $I$, not trapped diffusive domain walls. Our numerical results also indicate that the overlap distributions of PBC and FBC are likely to have the same thermodynamic limit. A rigorous proof of this would be interesting, yet challenging as FBC is not gauge related to PBC. Further investigations using larger system sizes and/or new statistics are needed to definitely understand the nature of the ordering of short-range Edwards-Anderson spin glasses.

\begin{acknowledgments}

We would like to thank D. L. Stein, M. A. Moore, J. Machta and H. G. Katzgraber for fruitful discussions and comments on an earlier version of the manuscript. W.W. acknowledges support from the National Science Foundation (Grant No.~DMR-1151387). The work is supported in part by the Office of the Director of National Intelligence (ODNI), Intelligence Advanced Research Projects Activity (IARPA), via MIT Lincoln Laboratory Air Force Contract No.~FA8721-05-C-0002. The views and conclusions contained herein are those of the authors and should not be interpreted as necessarily representing the official policies or endorsements, either expressed or implied, of ODNI, IARPA, or the U.S.~Government. The U.S.~Government is authorized to reproduce and distribute reprints for Governmental purpose notwithstanding any copyright annotation thereon. We thank Texas A\&M University for access to their Ada and Curie clusters.

\end{acknowledgments}

%\bibliographystyle{apsrevtitle}
%\bibliography{refs}

\end{document}